\documentclass[aps,prb,twocolumn, showpacs]{revtex4}
\usepackage{graphicx}
\usepackage{dcolumn}
\usepackage{amsmath}

\begin{document}

\title{Finite-size Effects in a Two-Dimensional Electron Gas with Rashba Spin-Orbit Interaction}
\author{C. P.  Moca$^{1,2}$ and D. C. Marinescu$^3$}
\affiliation{
$^1$Department of Physics, University of Oradea, 410087 Oradea, Romania \\
$^2$Institute of Physics, Technical University Budapest, Budapest, H-1521, Hungary\\
$^3$Department of Physics and Astronomy, Clemson University, 29634, Clemson
}

\date{\today}
\begin{abstract}
Within the Kubo formalism, we estimate the spin-Hall conductivity in
a two-dimensional electron gas with Rashba spin-orbit interaction
and study its variation as a function of disorder strength and
system size. The numerical algorithm employed in the calculation is
based on the direct numerical integration of the time-dependent
Schr\" odinger equation in a spin-dependent variant of the particle
source method. We find that the spin-precession length, $L_s$
controlled by the strength of the Rashba coupling, establishes the
critical lengthscale that marks the significant reduction of the
spin-Hall conductivity in bulk systems. In contrast, the electron
mean free path, inversely proportional to the strength of disorder,
appears to have only a minor effect.

\end{abstract}

\maketitle
\section {Introduction}
The physical phenomenon behind the spin-Hall effect
(SHE)\cite{Sinova,Murakami} in two-dimensional (2D) systems is the
flow of a pure spin current, spin-polarized in a transverse
direction, driven by a perpendicular electric field. Its existence
is conditioned by the presence of a spin-orbit interaction (SOI),
such as Rashba-Dresselhaus \cite{Rashba,Dresselhaus} in n-type two
dimensional systems  or the spin-split band structure in $p$-type
GaAs.\cite{sH}

If in clean samples the spin-Hall conductivity, $\sigma_{sH}$, was
predicted to have a universal, constant value of $e/8\pi$, in the
presence of disorder the resulting picture was less clear. It was
pointed out that, in 2D infinite systems, in the presence of short
range scatterers, the vertex corrections provided the exact
compensation to cancel the effect\cite{Inoue}. Moreover, an argument
was made that this cancellation occurs even for infinitesimal
disorder potentials\cite{Dimitrova}. These conclusions were
challenged by analytic\cite{Murakami2} and numerical
calculations\cite{Marinescu1,Nikolic, Shen} of the spin-Hall
conductivity in one and two-dimensional finite-size mesoscopic
samples, performed within the Landauer-B\" uttiker formalism, where
it was shown that the effect survives up to a critical disorder
strength.

Even though the robustness of the spin-Hall effect in the presence
of disorder seems to have been definitively confirmed by the
angle-resolved optical detection of spin polarization at opposite
edges of a two dimensional hole layer \cite{Wunderlich}, a better
understanding of the mechanism by which disorder and system size
affect spin transport in systems with spin-orbit interaction
warrants further investigation. We focus therefore on a study of the
interplay between the disorder strength, embodied in the electron
mean free path $l$, and the spin precession length $L_s$
proportional to the spin-orbit interaction, in determining the
spin-transport regime in finite size samples. Such an analysis is
especially relevant in two dimensions where, in the absence of any
additional interactions, the two lengths are independent of each and
along with the Fermi energy are the only relevant physical
parameters of the system.

The relationship between $l$ and $L_s$ and the system size, $L$,
 determines the existence of four distinctive transport
regimes. A semi-classical approximation is appropriate for $L_s \ll
L$, when the spin coherence is lost over the length of the sample,
while $L_s \gg L $ corresponds to a mesoscopic regime. When $l\gg L$
the electron propagation is ballistic, while for $l \ll L$ multiple
scattering events are assumed and the diffusive regime is present.

In the following analysis, we use the Kubo formula to estimate
 $\sigma_{sH}$ as a function of system size and disorder
in a two-dimensional electron system. The numerical formalism
adopted here represents an extension to the spin-Hall problem of the
particle-source method, developed by Tanaka\cite{Tanaka}. This
algorithm is based on the direct
 integration of the time dependent Schr\" odinger equation and allows
the calculation of the matrix elements of the Green's functions,
linear response functions, or any combinations of Green's function
and quantum operators in a very efficient way.

The main result of this paper is that the delimitation between the
mesoscopic and semiclassical regimes, as reflected by the rapid
decline of the spin-Hall conductivity, is established by $L_s$. For
system sizes smaller than $L_s$ the spin Hall conductivity increases
monotonically with the system size, while being weakly affected by
disorder. When $L\gg L_s$, $\sigma_{sH}$ decreases exponentially for
any amount of disorder in the system. This result supports the
conclusions of two previous reports by Sheng {\it et} al.
\cite{Sheng} and Nomura {\it et} al. \cite{Nomura} where it was
found that $\sigma_{sH}$ remains finite up to an, unspecified,
characteristic length scale and vanishes in the thermodynamic limit
for any small amounts of disorder in the system. Here, we identify
this length as being determined by the
 spin-precession length. Our results reflect no qualitative
modification of the overall behavior when the system evolves from
the ballistic to the diffusive regimes, crossover controlled by the
mean free path characteristic lengthscale. For a fixed Fermi energy
and $L_s$, the spin-Hall conductivity decrease monotonically with
disorder for any system size, as the system evolves from diffusive
to ballistic regime.

\section{Theoretical Framework}
\subsection{Model}

The single-particle Hamiltonian that describes the dynamics of an
electron of momentum $\mathbf{p}$ and effective mass $m^*$, is
written, in terms of the Pauli matrices $\sigma_{x,y}$ and the
Rashba coupling constant $\lambda$, as:
\begin{equation}
{\tilde H }= \frac{{\mathbf p}^2}{2 m^*} +\lambda \left( \sigma _x p_y -\sigma_y p_x\right ).
 \label{eq:hamiltonian}
\end{equation}
The exact diagonalization procedure that can be performed on the
Hamiltonian in the case of a clean system\cite{Sinova}, becomes
impossible when disorder is included in the form of an additional
random scattering term. It is therefore more convenient, for a
numerical analysis, to adopt the tight-binding approximation for the
many-body Hamiltonian, by employing a local orbital basis associated
with a virtual square $N\times N$ lattice, of constant $a_0$. In
this model, the many-body Hamiltonian, is:
\begin{eqnarray}
H &=&  \sum\limits_{i,\alpha}\varepsilon_i c_{i\alpha}^{\dagger} c_{i\alpha}
-t\sum\limits_{<i,j>,\alpha} c_{i\alpha}^{\dagger}c_{j\alpha}\label{eq:tight_binding_hamiltonian}\\
&+& V_R\sum\limits_{i, \delta_x, \delta_y} \left[\left( c_{i\uparrow}^{\dagger}c_{i+\delta_x \downarrow}-
c_{i\downarrow}^{\dagger}c_{i+\delta_x \uparrow}\right)\right . \nonumber \\
& &-i\left. \left( c_{i\uparrow}^{\dagger}c_{i+\delta_y \downarrow}+
c_{i\downarrow}^{\dagger}c_{i+\delta_y \uparrow}\right) \right]. \nonumber
\end{eqnarray}
In this expression, an electron with spin $\alpha$ at site $i$,
created by $c_{i\alpha}^{\dagger}$, is subjected to a random on-site
energy as in  the Anderson model for disorder, generated by a box
distribution $\varepsilon_i\in [-W/2, W/2]$. The electron transport
is described by a sequence of discrete hopping events.  Lateral
transport, without spin flip, to an adjacent site occurs with
probability $t = \hbar^2/2m^*a_0$, taken to be the unit of energy in
our calculation, as described by the second term in
Eq.~(\ref{eq:tight_binding_hamiltonian}). Propagation along the
diagonal sites, driven by the spin-orbit interaction, occurs with a
simultaneous spin-flip, as in the last term of
Eq.~(\ref{eq:tight_binding_hamiltonian}). The latter is most
important as it mixes the spin channels and leads to a finite
spin-Hall conductivity and spin accumulations at the edges of
sample. The Rashba coupling constant is renormalized by the lattice
constant to $V_R=\hbar \lambda / a_0$.

The Kubo formula for the spin-Hall conductivity is written as:
\begin{eqnarray}
\sigma _{sH}=\frac{1}{2}\, {\rm Tr}
\int \frac{d\varepsilon }{2\pi }
\left( -\frac{\partial f(\varepsilon )}{\partial \varepsilon }\right)
\left< {j}_x^z
\left[ {G}_R(\varepsilon )-{G}_A(\varepsilon )\right]
\right. \nonumber \\ \left. \times \,
{v}_y\, {G}_A(\varepsilon )
-{j}_x^z\, {G}_R(\varepsilon )\, {v}_y
\left[{G}_R(\varepsilon )-{G}_A(\varepsilon )\right]
\right>
\label{eq:kubo_spin_hall_conductivity}
\end{eqnarray}
The velocity operator is defined by the commutator: $i\, \hbar v_y =
\left[ y, H \right ]$, while for the spin current we adopt a
traditional expression\cite{Niu} given in terms of the
anticommutator between the velocity operator and the Pauli matrix
$\sigma_z$: $j_x^{z} =\hbar\left\{ \sigma_z, v_x \right\}/4$.
$G_{R/A}(\epsilon)$ represents the retarded/advanced Green's
function. In Eq. (\ref{eq:kubo_spin_hall_conductivity}) the
integration over the energy is restricted over the Fermi surface due
to the presence of $\left[ -\partial f(\varepsilon )/\partial
\varepsilon \right]$ factor.

In the tight-binding framework, the effect of disorder and
spin-orbit interaction strength on the spin Hall conductance was
investigated previously, using the Landauer-B\" uttiker
formalism\cite{Hankiewicz, Nikolic2}. As will be discussed in the
next section, in the present work we use a different approach for
computing the Green's function needed for the calculation of the
spin-Hall conductivity.

\subsection{Numerical algorithm}

Since the purpose of this investigation is an analysis of the
spin-Hall conductivity dependence on system size and disorder, we
will apply the Kubo formula to large size systems for different
values of the disorder potential W. The numerical algorithm that
underlies this calculation has been introduced in Ref.
\onlinecite{Tanaka} and represents an extension of the
particle-source method combined with tight-binding formalism.
 This method was first applied to the calculation
of the Green's function, density of states, conductivity
\cite{Iitaka1} and Hall conductivity\cite{Tanaka}. The main
advantage is that one can evaluate both the diagonal and
off-diagonal parts of the Green's function and their products with
other quantum operators with a low computing effort. In principle
the computing effort for computing the Green's function is $O(N^3)$,
(Hamiltonian is expressed as a $N\times N$ matrix) while within the
present algorithm only $O(N)$ computational effort is required for
the same calculation. Here, we briefly outline the main features of
the algorithm.

The central part of the method consists in solving the time
dependent Schr\" odinger equation with a single-frequency source
term:
\begin{equation}
i\hbar\frac{ d\, \left | \tilde{j}, t \right >}{d\, t} = H \left | \tilde{j}, t \right > + \left|j\right>\theta(t)
\exp^{-i(E+i\eta)t}\;, \label{eq:schr}
\end{equation}
where $\eta$ is a finite small value and $\theta$ is the step function.
The solution of the equation, with the initial
condition $\left | \tilde{j}, t=0 \right >=0$ becomes:
\begin{eqnarray}
\left | \tilde{j}, t \right > & = & -i\int_{0}^{t}dt'e^{-iH(t-t')}\left | {j}\right >
e^{-i(E+i\eta)t'}\\ \nonumber
& = &
\frac{1}{E+i\eta-H}[e^{-i(E+i\eta)t}-e^{-iHt)}\left | {j}\right >\;.
\end{eqnarray}
For sufficiently large time, one can then write the solution to the Schr\" odinger
equation in terms of the Green's function
acting on the "source" $\left|j\right>$ with the relative accuracy $\delta = e^{-\eta T}$, as:
\begin{equation}
\left |\tilde{j}, T\right >=G(E+i\eta)\left | {j}\right >e^{-i(E+i\eta)T}\;,
\end{equation}
leading to a Green's function operating on the ket $\left | {j}\right >$:
\begin{equation}
G(E+i\eta)\left | {j}\right >=\lim_{T\rightarrow\infty}\left |\tilde {j}, T\right >e^{i(E+i\eta)T}\;.
\end{equation}
The matrix element between states $\left<i\right|$ and $\left|j\right>$ is then obtained as:
\begin{equation}
\left<i\right|G(E+i\eta)\left|j\right> = \lim_{T\rightarrow \infty}\left< i \right|\left.\tilde {j}, T\right >\;.
\end{equation}
The matrix elements of a product including several Green's
functions and other operators are obtained by choosing a new
initial state, such as $\left|j'\right> = AG(E+i\eta)\left|j\right>$
in Eq.~(\ref{eq:schr}) and repeating the same procedure.

To calculate the matrix elements of the Green's function at many different energy values,
one solves Eq.~(\ref{eq:schr})
simultaneously for a source term with multiple frequencies,
 $\left|j\right>\left(\sum_l e^{-i(E_l+i\eta)t}\right)\theta(t)$. Following the algorithm outlined above, one obtains
 as an approximate solution, the ket
\begin{equation}
 \left|\tilde{j},T\right>\simeq\sum_l G(E_l+i\eta)\left|j\right>e^{-i(E_l+i\eta)T}.
 \end{equation}
 The matrix element of the Green's function between the states $\left < i\right |$ and $\left | j \right >$
for a giver energy is then easily obtained as
 \begin{eqnarray}
 G_{ij}(E_l' + i \eta ) & = & \left<i\right|G(E_l' + i \eta)\left|j\right> \nonumber\\
  & = & \frac{1}{T}\int_0^T dt'\left < i | \tilde{j},t'\right>e^{i(E_{l'}+i\eta)t'}\;,
 \end{eqnarray}
 where the terms involving transitions between different energies have been neglected with the relative accuracy
 $\delta = 1/T\Delta E$, with $\Delta E$ the minimum increment in the energies $E_l$.

To obtain the time dependent ket $\left |\tilde{j};T\right >$, a direct numerical integration of the Schr\" odinger equation can be
performed, as in the "leap-frog" algorithm\cite{Iitaka}. This is a second order, symmetrized differencing scheme, accurate
up to $(H\Delta t)^2$. In this form, Eq.~(\ref{eq:schr}) becomes:
\begin{eqnarray}
\left|\tilde{j};t+\Delta t\right>&=&-2i\Delta t H\left|\tilde{j};t\right> +\left|\tilde{j};t-\Delta t\right>\\ \nonumber
&-& 2i\Delta t \left|j\right>\sum_l e^{-i(E_l + i\eta)t}\theta (t)\;,
\end{eqnarray}
with a time step $\Delta t$ determined by $\Delta t = \beta/E_{max}$, where $E_{max}$ is the absolute value of the extreme
eigenvalue and $\beta$ is a parameter whose value is less than 1 in order for the solution to be stable\cite{Askar}.

Estimating the trace in Eq.~(\ref{eq:kubo_spin_hall_conductivity})
requires a suitable basis set, such as the local orbital basis. It
is more efficient however, to choose a randomized version of this
basis, described by a ket $\left| \phi \right > = \sum_{n=1}^{N}
\left | n \right > \exp(-i \, \phi_n)$, where $\left | n \right >$
are the tight-binding orbitals and $\phi_n$ are random numbers in
the $\left[0, 2\pi \right]$ interval. For a given operator $A$,
$\left < \phi\right| A\left | \phi \right >\simeq \sum_n \left < n
\right |  A\left| n \right >$ within the statistical errors of
$1/\sqrt{N}$.

\section{Results and Discussion}
\begin{figure}[h]
\centering
\includegraphics[width=3.0in]{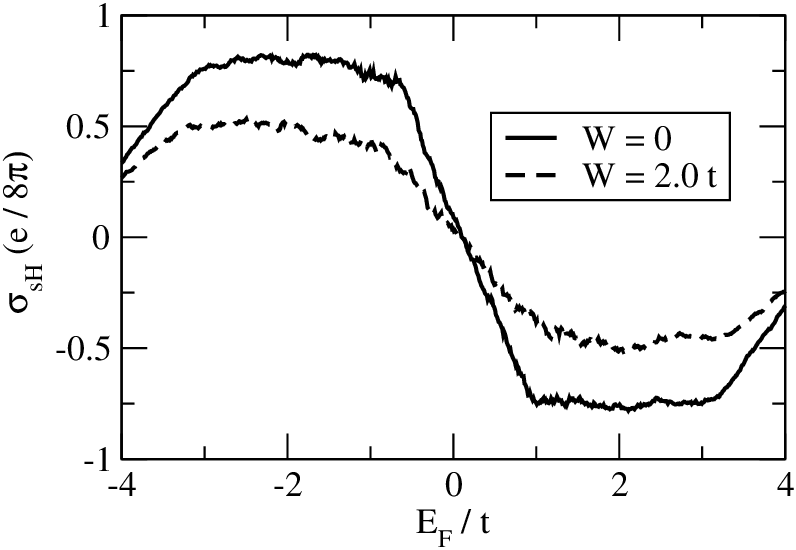}
\caption{ Spin Hall conductivity as function of Fermi Energy in units of $e/8\pi$, for
clean and disordered systems. System size is  30$\times$ 30. Spin-orbit
interaction strength is fixed to $V_R = 0.2\, t$. Average was done
over 2000 random frequencies.}
\label{fig:spin_hall_conductivity}
\end{figure}

In this section we show the results of our computation based on the
previously outlined algorithm. First,  we study the Fermi energy
dependence of the spin-Hall conductivity, presented in Fig.
\ref{fig:spin_hall_conductivity}. Random Fermi energies in the
interval [-4t,4t] were considered and we average over 2000 samples,
for each system size. For clean systems and for states in the band,
$\sigma_{sH}$ is close to 0.8  (in unit of $e/8\pi$), except at half
filling where due to electron-hole symmetry considerations it
vanishes. In the presence of disorder, the calculated value of
$\sigma_{sH}$ decreases, our results reproducing very well the known
behavior previously obtained in the Landauer-B\" uttiker
formalism\cite{Nikolic,Marinescu1} or by the analytical Kubo
formula\cite{Nomura}.

\begin{figure}[b]
\centering
\includegraphics[width=3.0in]{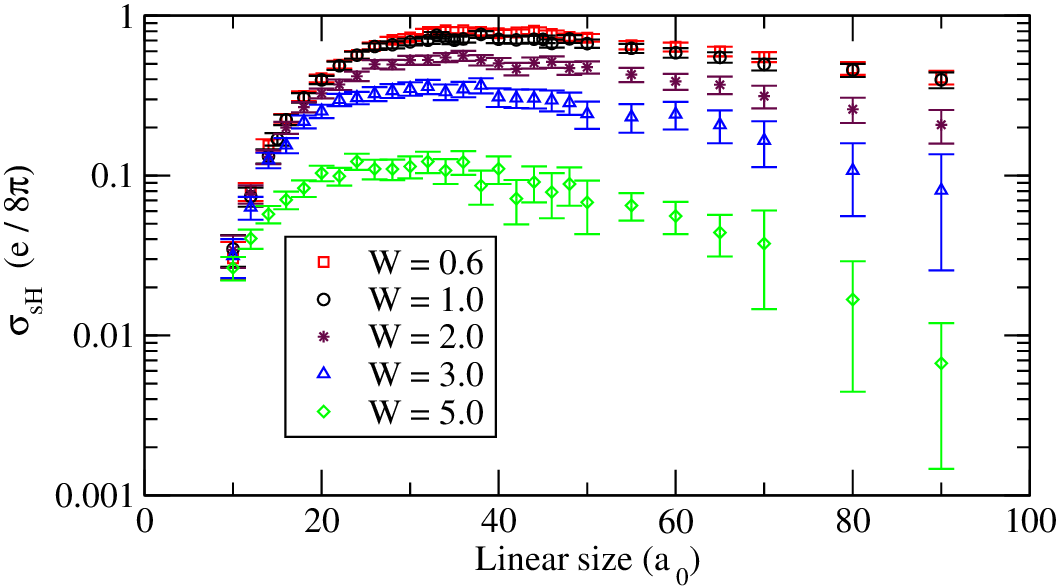}
\caption{Spin Hall conductivity as function of system size in units
of $e/8\pi$ represented on a logarithmic scale. Fermi energy is
fixed to $E_F = -3.0\;t $ and  the spin precession length is $L_s
\simeq 16 a_0$, corresponding to $V_R = 0.2\, t$. Mean free path
ranges from $l \simeq 45 a_0 $ for $W = 0.6\;$ down to a few lattice
constants when disorder increases. Averages are done over $10^3$
disorder samples and for each, 200 initial state vectors are
considered. } \label{fig:scaling_linear_size_all_range}
\end{figure}
The dependence of the spin-Hall conductivity on the system size is
shown in Fig. \ref{fig:scaling_linear_size_all_range}. One is
interested in finding whether the variation of $\sigma_{sH}$ is
dramatically changed by disorder and what is the length scale at
which this change occurs. For this, the two relevant parameters are
the electronic mean free path $l$ and the spin-precession length
$L_s$. In a quasi-classical approximation  $l = 12\hbar v_F a_0/(2
\pi N(E_F)\, W)^2$ , where $v_F$ is the Fermi velocity, while
$N(E_F)$ is the density of states at the Fermi energy measured from
the bottom of the band. The spin-precession length is defined in
terms of the Rasba coupling constant by $L_s = \pi t a_0/V_R $.

The electronic mean free path is the lengthscale that separates the
ballistic from the diffusive regimes, with a ballistic behavior for
system sizes smaller than $l$ and diffusive otherwise. We found that
the crossover between these two regime is smooth, without any
dramatic changes in the overall behavior of the spin-Hall
conductivity. The only observable effect is a decrease of the
spin-Hall conductivity when disorder increases. For example,  when
$W=0.6\; t$, $l\simeq45\;a_0$, while for $W =1.0\; t$, $l\simeq 16\;
a_0$, whereas, as can be seen in Fig.
\ref{fig:scaling_linear_size_all_range}, the behavior of the
spin-Hall conductivity remains unchanged. At the same time, for
system sizes below $L_s$, $\sigma_{sH}$ is always monotonically
increasing, reaches a plateau between $L_s$ and $2L_s$,  and then
decreases for large system sizes, being expected to vanish in the
thermodynamic limit, as in Refs.\cite{Sheng, Nomura}. The
spin-precession length, therefore, is the characteristic lengthscale
at which a crossover between the different regimes of the spin-Hall
conductivity is expected.
\begin{figure}[h]
\centering
\includegraphics[width=3.0in]{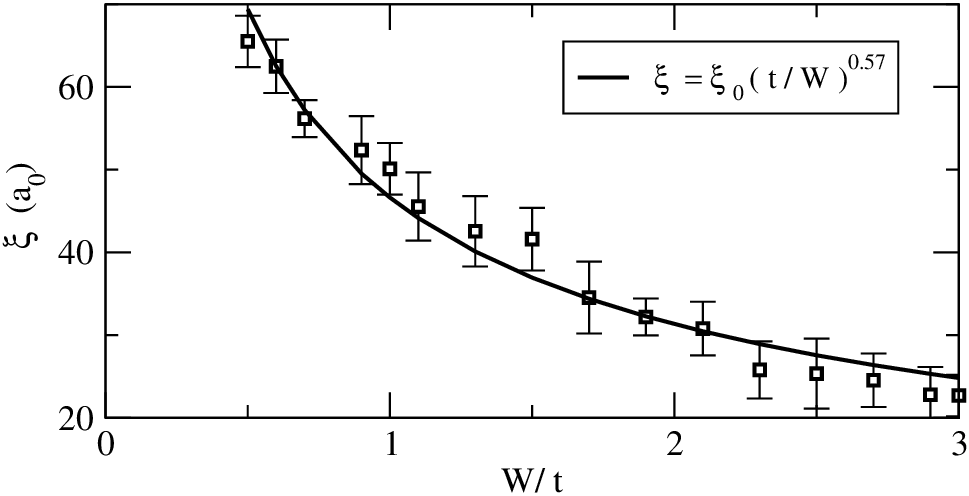}
\caption{ Disorder dependence
of the characteristic length scale $\xi$ for Fermi energy
$E_F = -3.0\, t$ and spin-orbit interaction strength $V_R = 0.2\, t $.}
\label{fig:coefficient}
\end{figure}
In the semiclassical regime, a scaling analysis is appropriate. We
find that, for a given Fermi energy, the size dependence of the
spin-Hall conductivity can be very well fitted with a exponential
function $\sigma_{sH}\simeq \exp(-L/\xi)$ where $\xi$ is a
characteristic length that depends on the disorder strength, being
divergent in a clean system. In Fig. \ref{fig:coefficient} we
present the dependence of this characteristic length as function of
disorder which follows a power law $\xi\simeq \xi_0 \left ( t /W
\right )^{0.57}$ behavior, with the best fit  $\xi_0\simeq 46\,
a_0$. We remark again that the system size where $\sigma_{sH}$
starts to decrease is strongly conditioned by $L_s$, rather than $l$
as can be inferred from the similarities between the behaviors drawn
for different values of the disorder for the same $L_s$. In all our
calculations the decrease in the spin-Hall conductivity starts for
system sizes $ L \simeq 2\, L_s$ regardless of the value of the
electronic mean free path.
\section {Conclusions}
In this work we study the effect of the spin-precession length scale
and of the electronic mean free path on the the spin-Hall
conductivity in different regimes, by adapting the particle-source
algorithm to spin transport in systems with SOI in the framework of
the  the tight-binding approximation. The dependence of $\sigma_{sH}$ on
the Fermi energy is also investigated.
 Our main finding is that the spin precession length is the critical length scale
for the spin-Hall behavior. For a system size smaller that $L_s$,
the spin-Hall conductivity increases even in the presence of
disorder, reaches a plateau between $L_s$ and $2\, L_s$ and then in
the semiclassical limit, when $L\gg L_s$ decreases exponentially. In
the thermodynamic limit, $\sigma_{sH}$ is zero for any amount of
disorder present in the system. We have also showed that the
electronic mean free path does not play a fundamental role in the
spin-Hall conductivity behavior.

\emph{ Acknowledgments.}
This research was supported by the Hungarian Grants OTKA
Nos. NF061726 and T046303, and by the Romanian Grant CNCSIS 1.97.

\end{document}